\documentclass[aps,preprint]{revtex4}
\usepackage[dvips]{graphics,graphicx}

\begin{document}
\title{Wannier-Stark states in  double-periodic lattices I: one-dimensional lattices}
\author{Dmitri.~N.~Maksimov$^{1}$, Evgeny~N.~Bulgakov$^{1}$, and Andrey~R.~Kolovsky$^{1,2}$}
\affiliation{$^1$Kirensky Institute of Physics, 660036 Krasnoyarsk, Russia}
\affiliation{$^2$Siberian Federal University, 660041 Krasnoyarsk, Russia}
\date{\today}

\begin{abstract}
We analyze the Wannier-Stark spectrum of a quantum particle in generic one-dimensional double-periodic  lattices. In the limit of weak static field the spectrum is shown  to be a  superposition of two Wannier-Stark ladders originated from two Bloch subbands. As the strength  of the field is increased, the spectrum rearranges itself into a single Wannier-Stark ladder. We derive  analytical expressions which describe the rearrangement employing the analogy between the Wannier-Stark problem and the  driven two-level system  in the strong-coupling regime.
\end{abstract}

\maketitle

\section{Introduction}
\label{sec1}

By definition, Wannier-Stark states (WS-states) are the eigenstates of a quantum particle in a periodic potential in the presence of a static field $F$. For a simple 1D lattice of the period $a$ the spectrum of WS-states is a ladder of energy levels with the level spacing $aF$, known as the Wannier-Stark ladder or the Wannier-Stark fan. The equidistant spectrum implies periodic dynamics of the particle which is nothing else as celebrated Bloch oscillations (BOs). If the lattice period is doubled,  BOs become a complicated process because of the Landau-Zener tunneling (LZ-tunneling) between two subbands that emerge from a single band due to the period doubling.  In the past decade BOs and LZ-tunneling in 1D double-periodic lattices has attracted much attention in cold atoms physics and photonics  thanks to applications to interferometric measurements and as a method for manipulating localized wave-packets \cite{Breid06,Breid07,Drei09,Kling10,Plot11}. The main question we address in this work is how the interband LZ-tunneling is  encoded in the properties of WS-states.  In fact,  since an arbitrary initial  quantum state of the system can be expanded over the basis of WS-states, they  provide an alternative approach for describing different dynamical phenomena,  including LZ-tunneling. The advantages of this alternative approach becomes especially transparent in two-dimensional systems which will be the subject of our subsequent paper \cite{preprint}. Thus the present work can be also  viewed as  a necessary step before proceeding with analysis of WS-states  in two-dimensional lattices.

The structure of the paper is as follows. In Sec.~\ref{sec2} we introduce the model -- the tight-binding Hamilltonian of a double-periodic lattice and perform preliminary analysis  of the Wannier-Stark spectrum (WS-spectrum). This analysis reveals two different regions in the parameter space -- the cases of weak and strong fields -- which are analyzed in detail in  Sec.~\ref{sec3}.  We obtain asymptotic expressions for the WS-spectrum in the limit   $F\rightarrow\infty$ and $F\rightarrow0$ and discuss two analytical methods that describe  this spectrum  for intermediate $F$. Finally, in Sec.~\ref{sec4} we analyze the system beyond  the tight-binding approximation to see  effects which are neglected by this. The main results are summarized in the concluding Sec.~\ref{sec5}.

\section{The model}
\label{sec2}

\begin{figure}
\center
\includegraphics[width=10cm,clip]{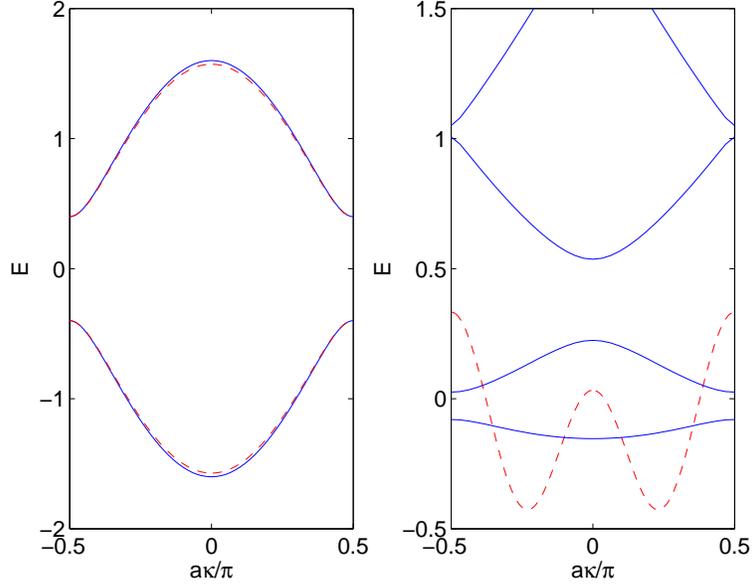}
\caption{(Color online) Left panel: Bloch bands of the lattice (i) for $J= 0.76$ and $\delta=0.4$ (dashed line) and the lattice (ii) for $J_1 = 1$ and $J_2=0.6$ (solid line). Right panel: Bloch spectrum of the system (\ref{a16}) for $(V_0,V_1,V_2)=(-0.117 -0.15 0.3)$. The dashed line shows the double-periodic potential $V(x)$ for the specified parameters. }
\label{fig1}
\end{figure}

Within the tight-binding approximation an arbitrary double-periodic lattice is characterized by four parameters -- alternating tunneling elements $J_1$ and $J_2$, alternating on-site energies $\pm\delta$, and the Stark energy $F$ (we set the distance $a$ between the nearest sites to unity). For $F=0$ the spectrum of the system consists of two Bloch bands,
\begin{equation}
\label{1}
E_\pm(\kappa)=\pm\sqrt{\delta^2+J_1^2+J_2^2+2J_1J_2\cos(2\kappa)} \;,
\end{equation}
where $\kappa$ is the quasimomentum defined in the reduced Brillouin zone, $-\pi/2\le\kappa<\pi/2$. In what follows we shall be mainly concerned with two cases:  (i) $J_1=J_2\equiv J$ yet $\delta\ne0$; (ii) $\delta=0$ yet $J_1\ne J_2$. These two lattices can have almost indistinguishable Bloch spectrum, see Fig.~\ref{fig1}(a), however, their Bloch states are profoundly different. In fact, the Bloch states of the lattice (ii), known in the solid state physics as the SSH-lattice \cite{Su79},  possess nontrivial topological properties reflected in the quantized Zak phase \cite{Zak89}. On the contrary, the case (i) corresponds to a topologically trivial lattice. We mention in passing that recently the Zak phase has been measured in cold-atom implementation of the SSH-lattice \cite{Atal13}.

\begin{figure}
\center
\includegraphics[width=10cm,clip]{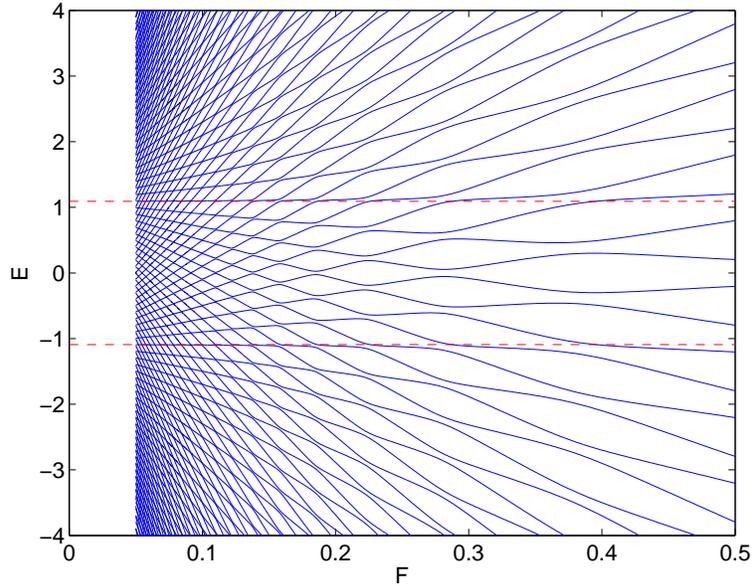}
\caption{(Color online) Wannier-Stark spectrum for the lattice (ii) as the function of $F$. The dashed lines correspond to Eq.~(\ref{3}) with $n=0$.}
\label{fig2}
\end{figure}

\begin{figure}
\center
\includegraphics[width=10cm,clip]{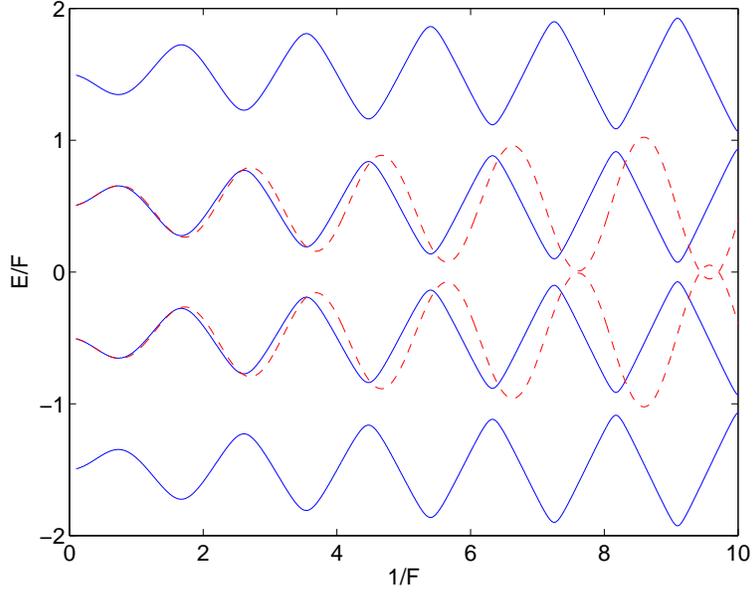}
\caption{(Color online) Scaled spectrum as the function of $1/F$. The dashed lines correspond to Eq.~(\ref{24}).}
\label{fig3}
\end{figure}

If $F\ne0$ the continuous Bloch spectrum (\ref{1}) transforms into the discrete WS-spectrum. For the sake of preliminary analysis  we calculate the spectrum using the straightforward diagonalization of the Hamiltonian matrix. Denoting the occupation probabilities for sites $A$ and $B$ by $\psi_l^{A,B}$ (here index $l$ labels  elementary cells consisting of two sites), we have the stationary Schr\"odingier equation for the tilted double-periodic lattice in the form
\begin{eqnarray}
\nonumber
2F(l-1/4) \psi_l^{A}-\delta \psi_l^{A} - J_2 \psi_l^{B} - J_1\psi_{l-1}^{B}  =E\psi_l^{A} \;,\\
\label{2}
2F(l+1/4) \psi_l^{B}+\delta \psi_l^{B} - J_2 \psi_l^{A} - J_1 \psi_{l+1}^{A} =E\psi_l^{B} \;,
\end{eqnarray}
where the Stark term corresponds to the potential energy $U(x)=F(x-x_0)$ with $x_0$ chosen in the middle between $A$ and $B$ sites. The solid lines in Fig.~\ref{fig2} show numerical solution of Eq.~(\ref{2}) as the function of $F$ for the SSH-lattice. It is seen that the spectrum consists of two Wannier-Stark fans that are associated with two Bloch bands in Fig.~\ref{fig1}(a). In the region of large $F$ the ladders strongly affect each other  that is reflected in pronounced avoided crossings. The gap of the avoided crossings, however, progressively decreases if $F\rightarrow0$. This is clearly seen in Fig.~\ref{fig3} where we scale the spectrum according to the ladder spacing $F$. Thus in the limit of small $F$ we have
\begin{equation}
\label{3}
E_{n,\pm} \approx \pm C + 2Fn, \ \ n=0,\pm 1,\ldots \;,
\end{equation}
where the constant $C$ will be specified later on in Sec.~\ref{sec3b}. It is also seen in Fig.~\ref{fig3} that in the opposite limit of large $F$ two Wannier-Stark ladders merge into one ladder with the level spacing $F$, i.e.,
\begin{equation}
\label{3b}
E_n \approx  F(n+1/2), \ \ n=0,\pm 1,\ldots \;.
\end{equation}

To calculate the spectrum using Eq.~(\ref{2}) we truncate infinite system of equations to a finite system which results in numerical errors. In the next section we describe an approach which is free from this drawback and, what is more important, opens a way for finding analytical solutions.

\section{Floquet operator approach}
\label{sec3}

To approach Eq.~(\ref{2}) analytically we  introduce the generating functions
\begin{equation}
\label{4a}
Y^{A,B}(\theta)=(2\pi)^{-1/2} \sum_{l=-\infty}^\infty \psi_l^{A,B}\exp(il\theta) \;.
\end{equation}
This reduces   Eq.~(\ref{2}) to the system of two ordinary differential equations:
\begin{equation}
\label{4}
i2F\frac{{\rm d {\bf Y}(\theta)}}{{\rm d}\theta}=G(\theta){\bf Y}(\theta),
\end{equation}
where ${\bf Y}(\theta)={[Y^{A}(\theta),Y^{B}(\theta)]}$ and $2\times2$ matrix  $G(\theta)$ is given by
\begin{equation}
\label{4_1}
G(\theta)=\left(
\begin{array}{cc}
E+F/2+\delta&J_2+J_1\exp(-i\theta)\\
J_2+J_1\exp(i\theta)&E-F/2-\delta
\end{array} \right).
\end{equation}
%
%
Since  $Y^{A,B}(\theta)$ are by definition periodic functions of $\theta$  we are only interested in periodic solutions of  Eq.~(\ref{4}). This gives the quantization rule for the energy $E$ entering Eq.~(\ref{4}). The periodicity of solutions implies that eigenvalues of the evolution (Floquet) operator
\begin{equation}
\label{5}
U=\widehat{\exp}\left[-\frac{i}{2F}\int_0^{2\pi} G(\theta){\rm d}\theta \right]
\end{equation}
must be unity. Numerically, we can use this fact to find the Wannier-Stark spectrum exactly, i.e., without using the truncation procedure. In more detail, first we calculate (\ref{5}) for a trial energy $E=0$ and diagonalize it. This provides two complex eigenvalues $\lambda_1$ and $\lambda_2=\lambda_1^*$. Then the positions of energy levels in Fig.~\ref{fig2} or Fig.~\ref{fig3} are found from the equation
\begin{equation}
\label{6}
\exp\left(-i\frac{\pi E}{F}\right)=\lambda_{1,2} \;.
\end{equation}
%

Unfortunately, Eq.~(\ref{4}) has no analytical solution in the closed form which would be valid in the whole parameter space. Nevertheless, we can obtain analytical solution in the case of weak fields and separately in the case of strong fields. A quantity, which distinguishes these two cases, is obviously the size of the energy gap separating two Bloch subbands as compared to the Stark energy. In terms of Bloch dynamics it distinguishes the  regime of  negligible interband LZ-tunneling from that where the tunneling is the main effect. We begin with the case of strong fields.

\subsection{Strong fields}
\label{sec3a}

As it was already mentioned in Sec.~\ref{sec2}, in the limit of large $F$ two ladders are strongly coupled that leads to almost equidistant spectrum with the level spacing $F$.  The parameters, which quantify the strength of coupling, are
\begin{equation}
\label{11}
\epsilon_{1}=(J_2-J_1)/F \;,
\end{equation}
if $\delta=0$, and
\begin{equation}
\label{11a}
\epsilon_{2}=\delta/F \;,
\end{equation}
if $J_1=J_2$ but $\delta\ne 0$. The maximal coupling corresponds to $\epsilon_{1}=0$ ($\epsilon_{2}=0$) that is reached either by taking the limit $F\rightarrow\infty$ or by closing the energy gap between Bloch subbands. In terms of Eq.~(\ref{4}) this corresponds to the trivial solutions
\begin{equation}
\nonumber
{\bf Y}_{-}(\theta)= \frac{1}{\sqrt{2\pi}}\left(
\begin{array}{c}
e^{in\theta}\\0
\end{array} \right) \;, \quad
{\bf Y}_{+}(\theta)= \frac{1}{\sqrt{2\pi}}\left(
\begin{array}{c}
0\\e^{in\theta}
\end{array}
\right) \;,
\end{equation}
with the energies $E_{n,+}=F(2n+1/2)$ and $E_{n,-}=F(2n-1/2)$, respectively. To find the periodic solutions of Eq.~(\ref{4}) for finite  $\epsilon_{1}$ or/and $\epsilon_{2}$ we use (and compare) two different methods: Wu-Yang iterative approach from the theory of periodically driven two-level systems \cite{Wu07} and a perturbative approach based on the Bogoliubov-Mitropolskii averaging  technique from the theory of classical dynamical systems \cite{Mitr71}.

\subsubsection{Wu-Yang iterative approach}

\begin{figure}
\center
\includegraphics[width=10cm,clip]{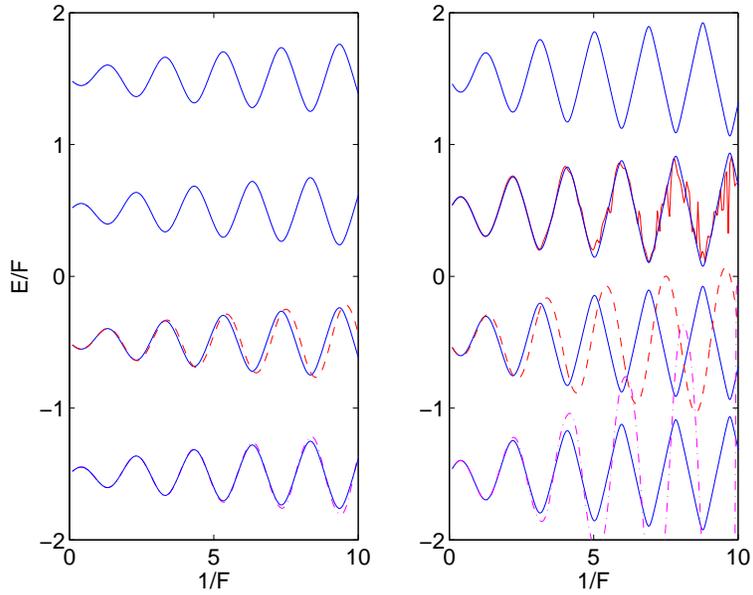}
\caption{(Color online) Scaled spectrum for the lattice (i) with $J_1=J_2=0.76$ and $\delta=0.2$, left panel, and $\delta=0.4$, right panel. The dashed and dash-dotted lines are the first and third order approximations according to Eq.~(\ref{24}), the broken line corresponds to Eq.~(\ref{23}). }
\label{fig4}
\end{figure}

Let us consider the lattice (i), i.e., $J_1=J_2\equiv J$ and $\delta\ne0$. After the substitution
\begin{equation}
\label{13}
Y^{A}=\tilde{Y}^A \exp(-iE\theta/2F-i\theta/4) \;,\quad  Y^B=\tilde{Y}^B \exp(-iE\theta/2F+i\theta/4)
\end{equation}
and $t=\theta/2$ Eq.~(\ref{4}) takes the form of  Schr\"odinger equation for a periodically driven two-level system:
\begin{equation}
\label{14a}
i\frac{{\rm d}}{{\rm d} t}\left(
\begin{array}{c}
\tilde{Y}^A\\ \tilde{Y}^B
\end{array} \right)=\left(
\begin{array}{cc}
\epsilon_2&\Omega\cos t\\
\Omega\cos t &-\epsilon_2
\end{array} \right) \left(
\begin{array}{c}
\tilde{Y}^A\\ \tilde{Y}^B
\end{array}\right) \;,
\end{equation}
where $\Omega=\frac{2J}{F}$ plays the role of the Rabi frequency. Since we are interested in the limit $\epsilon_2\ll \Omega$ we are in the so-called strong-coupling regime where the common rotating-wave approximation is not justified. This regime has attracted much attention in quantum optics  -- we shall follow the above cited work \cite{Wu07} which reports recent progress in the strong-coupling problem. Essentially the method provides an approximate expression for the evolution operator $U(t)$,
\begin{equation}
\label{21}
\tilde{{\bf Y}}(t)=U(t)\tilde{{\bf Y}}(0) \;,
\end{equation}
which is given in the Appendix.  To satisfy the periodicity of the solutions, Eq.~(\ref{21}) should be complemented with the `boundary conditions'
\begin{equation}
\label{22}
\tilde{Y}^A(\pi)=\exp\left(-i\frac{E\pi}{F}\right) \tilde{Y}^A(0) \;,\quad \tilde{Y}^B(\pi)=-\exp\left(-i\frac{E\pi}{F}\right) \tilde{Y}^B(0) \;.
\end{equation}
This yields the spectrum
\begin{eqnarray}
\label{23}
E_{n,\pm}=F(2n\pm1/2) \pm \frac{F}{\pi}\arcsin\left(\frac{U_{11}(\pi)-U_{22}(\pi)}{2i}\right)  \;.
\end{eqnarray}
Expanding Eq.~(\ref{23}) in the parameter $\epsilon\equiv\epsilon_2$ up to the forth order, we have
\begin{eqnarray}
\label{24}
E_{n,\pm}=F(2n\pm1/2) \pm \epsilon \Pi_1(F) \pm \epsilon^3\Pi_3(F) \;,
\end{eqnarray}
where
\begin{equation}
\nonumber
\Pi_1(F)=F{\cal J}_0\left(\frac{4J}{F}\right)
\end{equation}
and
\begin{equation}
\nonumber
\Pi_3(F)=\frac{2F}{\pi}\int\limits_{0}^{\pi}dt{\left[{\cal I}\left(t,\frac{4J}{F}\right)-\frac{1}{2}{\cal I}\left(\pi,\frac{4J}{F}\right) \right]}^2 \;.
\cos\left(\frac{4J}{F}\sin t\right),
\end{equation}
In the last two equations ${\cal J}_0(z)$ is the Bessel function of the first kind and
\begin{equation}
\nonumber
{\cal I}(t,z)=\int\limits_{0}^{t}dx\sin(z\sin x) \;.
\end{equation}
The accuracy of the asymptotic Eq.~(\ref{24}) is illustrated in Fig.~\ref{fig4}. In this figure the solid blue lines are the exact spectrum calculated by using Eqs.~(\ref{5}-\ref{6}), the dashed red lines -- the first order corrections to the zero order result, and the dash-dotted magenta line -- the third  order corrections. It is seen  in Fig.~\ref{fig4}(a) that the first order result systematically shifts positions of the avoided crossings. This is corrected by  the third order term in Eq.~(\ref{24}) -- now the avoided crossings  (more exactly, remnants of the avoided crossings) appear at the right position. Unfortunately, applicability of  Eq.~(\ref{24}) is restricted to small $\epsilon$ and if we increase $\delta$ this automatically decreases the validity interval on $F$, see Fig.~\ref{fig4}(b). In this figure we also depict the result according to Eq.~(\ref{23}).  It is seen  in Fig.~\ref{fig4}(b) that Eq.~(\ref{23}) removes the divergence of Eq.~(\ref{24}) but introduces unphysical oscillations.

\subsubsection{Bogoliubov-Mitropolskii averaging technique}

Next we discuss the perturbative approach based on the Bogoliubov-Mitropolskii averaging technique \cite{Mitr71}. In this subsection we shall consider the general case $\delta\ne0$ and $J_1\ne J_2$. Let us rewrite Eq.~(\ref{4}) in terms of parameters (\ref{11}) and (\ref{11a}) . This is done by using two substitutions. The first substitution defined in Eq.~(\ref{13}) results in the equation
\begin{equation}
\label{14}
i2F\frac{{\rm d}}{{\rm d}\theta}\left(
\begin{array}{c}
\tilde{Y}^A\\ \tilde{Y}^B
\end{array} \right)=\left(
\begin{array}{cc}
\delta&g(\theta)\\
g^*(\theta)&-\delta
\end{array} \right) \left(
\begin{array}{c}
\tilde{Y}^A\\ \tilde{Y}^B
\end{array}\right) \;,
\end{equation}
where $g(\theta)=J_1\exp(i\theta/2)+J_2\exp(-i\theta/2)$. The second substitution is
\begin{equation}
\label{15}
u=(\tilde{Y}^A+\tilde{Y}^B)\exp\left(-\frac{i}{2F}\int {\rm Re}[g] {\rm d}\theta \right) \;,\quad
v=(\tilde{Y}^A-\tilde{Y}^B)\exp\left(+\frac{i}{2F}\int {\rm Re}[g] {\rm d}\theta \right) \;.
\end{equation}
This gives
\begin{equation}
\label{16}
i\frac{{\rm d}}{{\rm d}\theta}\left(
\begin{array}{c}
u\\ v
\end{array} \right)=\left(
\begin{array}{cc}
0&f(\theta)\\
f^*(\theta)&0
\end{array} \right) \left(
\begin{array}{c}
u \\ v
\end{array} \right)  \;,
\end{equation}
where
\begin{equation}
\nonumber
f(\theta)=\left[\epsilon_{2}+i\epsilon_{1}\sin\left(\frac{\theta}{2}\right)\right]
\exp\left[i\frac{2(J_1+J_2)}{F}\sin\left(\frac{\theta}{2}\right)\right]\;.
\end{equation}
Since the function $f(\theta)$ is proportional to small parameters, Eq.~(\ref{16}) can be treated by the Bogoliubov-Mitropolskii perturbative approach.

The first oder of the  Bogoliubov-Mitropolskii theory amounts to replacing the function $f(\theta)$ in Eq.~(\ref{16}) by its mean value
\begin{equation}
\label{18}
\bar{f} = \frac{\delta}{F} {\cal J}_0\left(\frac{2(J_1+J_2)}{F}\right)
+\frac{J_1-J_2}{F} {\cal J}_1\left(\frac{2(J_1+J_2)}{F}\right)\;,
\end{equation}
where ${\cal J}_0(z)$ and ${\cal J}_1(z)$ are the Bessel functions of the first kind. After the above substitution Eq.~(\ref{16}) is trivially solved, providing two independent solutions. Next, using the substitutions  (\ref{13}) and (\ref{15}) in the reverse order we find two independent approximate solutions of Eq.~(\ref{4}). Finally, requiring that these solutions are periodic in $\theta$ we obtain corrections to the equidistant spectrum:
\begin{equation}
\label{19}
E_{n,\pm}=F\left(2n\pm \frac{1}{2} \pm \bar{f}(F)\right)  \;.
\end{equation}
If $J_1=J_2$ the above coincides with the first order corrections obtained in the previous subsection. If $J_1\ne J_2$, i.e. for the lattice (ii), the approximate solution (\ref{19}) is depicted in Fig.~\ref{fig3} by the red dashed lines. Notice a different asymptotic behavior at $1/F\rightarrow 0$ as compared to the lattice (i).

Comparing two methods used in this work we conclude that both methods give a tractable analytical expression only in the first order over $1/F$. Furthermore, when restricted to the first order, the Bogoliubov-Mitropolskii technique is simpler and more universal than the Wu-Yang approach.

\subsection{Weak field regime}
\label{sec3b}

\subsubsection{Geometric phase and asymptotic solution}

\begin{figure}
\center
\includegraphics[width=10cm,clip]{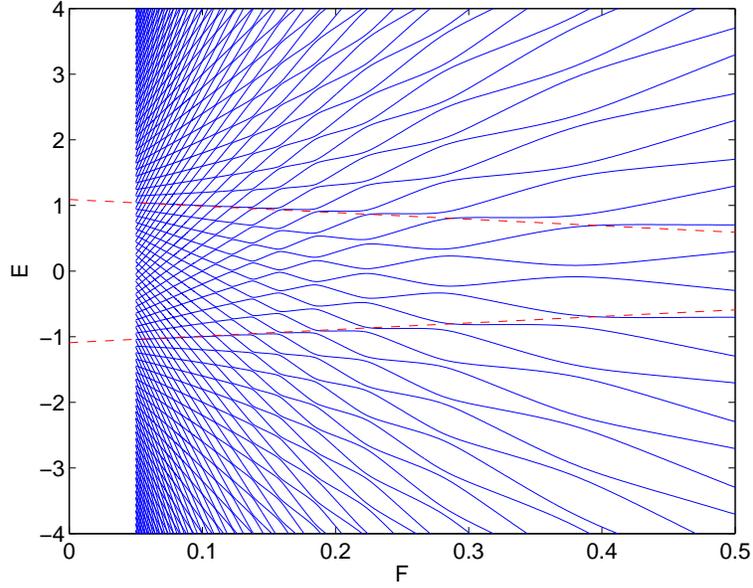}
\caption{(Color online) The same as in Fig.~\ref{fig2} yet for different dimerization $(J_1,J_2)=(0.6,1)$.}
\label{fig5}
\end{figure}

We proceed with the weak field limit where we shall focus on the lattice (ii). Assuming $F$ is out of vicinity of the avoided crossings, the periodic solution of Eq.~(\ref{4}) can be found by using the adiabatic theorem. It expresses the function ${\bf Y}(\theta)=[Y^A(\theta),Y^B(\theta)]^T$ in terms of instantaneous eigenfunctions ${\bf y}_{1,2}(\theta)$ of the $2\times2$ matrix $G(\theta)$ Eq. (\ref{4_1}),
\begin{equation}
\nonumber
G(\theta){\bf y}_{\pm}(\theta)={\cal E}_{\pm}(\theta){\bf y}_{\pm}(\theta) \;.
\end{equation}
We have
\begin{equation}
\label{a8}
{\bf Y}_{\pm}(\theta)=e^{-i\Phi_d(\theta)}e^{-i\Phi_g(\theta)} {\bf y}_{\pm}(\theta) \;,
\end{equation}
where
\begin{equation}
\nonumber
\Phi_d(\theta)=\frac{1}{2F}\int_0^\theta {\cal E}_{\pm}(\theta') {\rm d}\theta'
\;,\quad
\Phi_g(\theta)=i\int_0^\theta {\bf y}_{\pm}^T(\theta')\frac{{\rm d}}{{\rm d}\theta'} {\bf y}_{\pm}(\theta') {\rm d}\theta'
\end{equation}
are the dynamical and geometric phases, respectively. It is easy to prove that the eigenvalues ${\cal E}_{\pm}(\theta)$ are given by
\begin{equation}
\label{a10}
{\cal E}_\pm(\theta)=E + \tilde{{\cal E}}_\pm(\theta)   \;,\quad
\tilde{{\cal E}}_\pm(\theta) = \pm \sqrt{\left(\delta+\frac{F}{2}\right)^2+J_1^2+J_2^2+2J_1J_2\cos(\theta)} \;.
\end{equation}
To insure periodicity the solution (\ref{a8}) must satisfy the condition $\Phi_d(2\pi)+\Phi_g(2\pi)=2\pi n$ where $n$ is an integer. This results in the spectrum
\begin{equation}
\label{a11}
E_{n,\pm}=C_\pm+2F(n+c_\pm) \;,
\end{equation}
where
\begin{equation}
\label{a12}
C_\pm=\frac{1}{2\pi}\int_0^{2\pi} \tilde{{\cal E}}_{\pm}(\theta)   {\rm d}\theta \;,\quad
c_\pm=\frac{i}{2\pi}\int_0^{2\pi} {\bf y}_{\pm}^T(\theta)\frac{{\rm d}}{{\rm d}\theta} {\bf y}_{\pm}(\theta) {\rm d}\theta'.
\end{equation}
Comparing Eq.~(\ref{a10}) with the Bloch dispersion relation (\ref{1}) we conclude that in the limit of small $F$ the constants $C_\pm$ are given by the mean energies of the Bloch subbands,
\begin{equation}
\label{a13}
\lim_{F\rightarrow0} C_\pm=\frac{1}{\pi}\int_{-\pi/2}^{\pi/2} E_{\pm}(\kappa)   {\rm d} \kappa \;,\quad C_{+}=-C_{-}\equiv  C \;,
\end{equation}
and the constants $c_\pm$ are the Zak phases of these bands. For the example considered in Fig.~\ref{fig2} $c_\pm=0$ and, hence, we recover Eq.~(\ref{3}). However, for the alternative dimerization of the SSH-lattice $J_2>J_1$, one has $c_\pm=\pm 1/2$ and Eq.~(\ref{3}) must be corrected as $E_{n,\pm}=\pm C+2F(n\pm1/2)$, see Fig.~\ref{fig5}. As it was already mentioned in Sec.~\ref{sec2},  the SSH-lattice is a topological system, i.e., its geometric phase is insensitive to variation of the tunneling rates and  depends only on the dimerization.  This result does not hold in topologically trivial case $\delta\ne0$ where the constants $c_\pm$ in Eq.~(\ref{a11}) do depend on the lattice parameters and, hence, differ from both 0 or $\pm1/2$.

The accuracy of the adiabatic equation (\ref{a11}) can be improved by including the second order corrections that are proportional to $F^2$. Using the analogy with spin dynamics and adopting results of Ref.~\cite{Blio08} to the considered problem, the proportionality coefficient in front of $F^2$ is found as
\begin{equation}
\nonumber
D=\frac{(J_1+J_2)^2(J_1-J_2)^2}{32}\frac{1}{2\pi}\int_0^{2\pi}
\left[(J_1+J_2)^2\cos^2(\theta/2)+(J_1-J_2)^2\sin^2(\theta/2)\right]^{-5/2} {\rm d}\theta \;.
\end{equation}

\subsubsection{Avoided crossings and resonant tunneling}

\begin{figure}
\center
\includegraphics[width=10cm,clip]{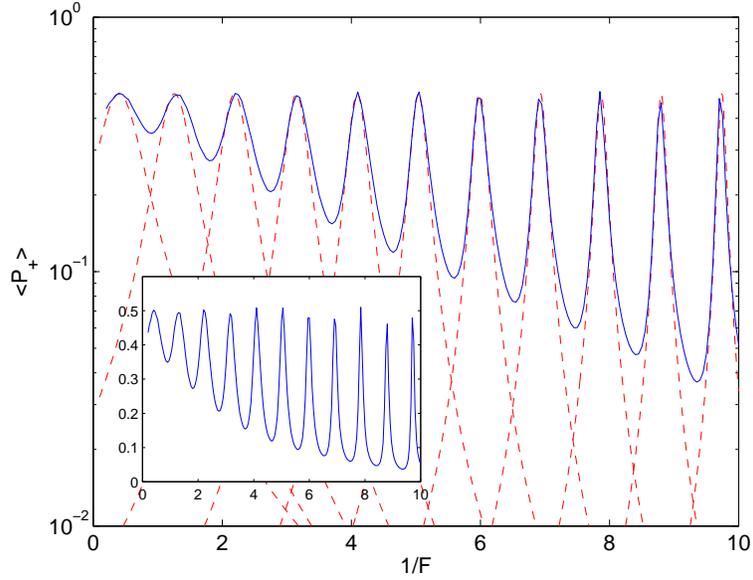}
\caption{(Color online) Time-averaged population of the upper band for the lattice (i) as the function of $1/F$ in the logarithmic (main panel) and linear (inset) scales .  Initially the whole lower band is populated. Average over 20 Bloch periods. (For infinite number of Bloch periods the hight of all peaks is $1/2$ exactly.) The dashed lines are approximations of the resonance peaks by the Lorentzian (\ref{add}).}
\label{fig8}
\end{figure}

One important point requiring special attention is that the adiabatic equation (\ref{a11}) breaks down at the level crossings. Here the level crossings should be replaced with avoided crossings with the gap $\Delta E$. Drawing analogy with the driven two-level system,  where the avoided crossings are associated with multiphoton resonances \cite{Krai80}, we have
\begin{equation}
\nonumber
\frac{\Delta E}{F}=\frac{2}{\pi}\exp\left(-\frac{1}{F}\int_0^{\theta_0}\sqrt{1 - \frac{2J_1J_2}{J_1^2+J_2^2}\cosh(\theta)} {\rm d}\theta \right) \;,\quad
\frac{2J_1J_2}{J_1^2+J_2^2}\cosh(\theta_0)=1 \;.
\end{equation}
%
It is easy to show that the avoided crossings between Wannier-Stark levels describe the so-called phenomenon of the resonant LZ-tunneling \cite{Plot11}. This phenomenon can be detected by analyzing
 population dynamics of the Bloch subbands. In fact, let us assume that initially only the lower band is populated and consider the mean (i.e., time-averaged) occupation of the upper band $P_+=\langle P_+(t) \rangle$ as the function of the static field. For the lattice (i) the result of this experiment is depicted in Fig.~\ref{fig8}. This figure should be compared with Fig.~\ref{fig4}(b). It is seen that positions of the resonance peaks coincide with positions of the avoided crossings in Fig.~\ref{fig4}(b) while the widths of peaks are determined by the gaps  $\Delta E_j$, so that locally one has
\begin{equation}
\label{add}
P_+(z)=0.5\frac{(\Delta E_j/2)^2}{(\Delta E_j/2)^2 + (z-z_j)^2} \;,\quad z=\frac{1}{F} \;.
\end{equation}

\begin{figure}
\center
\includegraphics[width=10cm,clip]{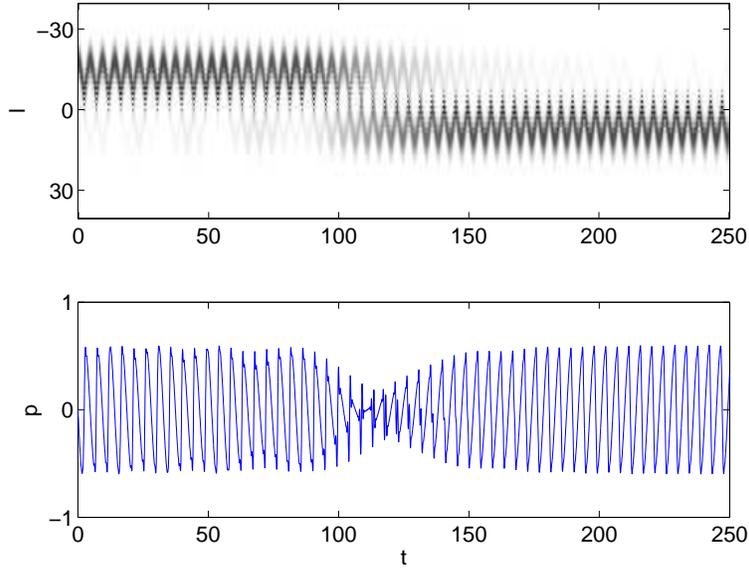}
\caption{(Color online) Gray-scaled image of the wave-packet (upper panel) and the mean momentum (lower panel) as the functions of time. Within the considered time interval static force is linearly increased from $F=1/9.4$ to $F=1/8.7$. The other parameters are $(J_1,J_2)=(1,0.6)$ and $\delta=0$. The time is measured in units of $T_J=2\pi/J_1$.}
\label{fig6}
\end{figure}

An interesting dynamical manifestation of the resonant tunneling is a possibility of transferring quantum particle from the lower Bloch subband to the upper subband and vice versa. Assume that $F$ is out of a given avoided crossing and the initial state of the system belongs to the lower subband. Then the particle performs BOs in the lower subband with negligible  LZ-tunneling to the upper subband. If we now adiabatically change $F$ to pass through the avoided crossing, the particl  will perform BOs in the upper subband. This dynamics is  illustrated in Fig.~\ref{fig6} which shows  BOs of a  localized packet. In this simulation we linearly change $F$ in the interval $8.7<1/F<9.4$ which  contains one avoided crossing at  $1/F\approx 9$ [see Fig.~\ref{fig3}].

\section{Beyond the tight-binding approximation}
\label{sec4}

In this section we discuss the cold-atom implementation of double-periodic lattices considered in the previous sections. After an appropriate rescaling, the dimensionless Hamiltonian of the system reads
\begin{equation}
\label{a16}
\widehat{H}=-\frac{1}{2}\frac{{\rm d}^2}{{\rm d} x}^2 +V(x)+Fx \; \quad V(x)=V_0+V_1\cos(2\pi x+\phi_1)+V_2\cos(4\pi x+\phi_2) \;,
\end{equation}
where $V_1$ and $V_2$ are proportional to intensities of two standing laser waves forming the optical lattice \cite{Kling10,Atal13}. For numerical purpose we introduce  additional parameter $V_0$ in the Hamiltonian (\ref{a16}) to shift the energy axis. Varying the parameters of the double-periodic potential $V(x)$ one can realize different values of the hopping matrix elements $J_1$ and $J_2$ and on-site energy $\pm\delta$ in the tight-binding model. We set $\phi_1=\phi_2=0$, that insures $\delta=0$, and  consider $|V_1|<|V_2|$, that ensures $J_2< J_1$. The Bloch spectrum of the system (\ref{a16}) together with the chosen potential $V(x)$ are shown in Fig.~\ref{fig1}(b).

\begin{figure}
\center
\includegraphics[width=10cm,clip]{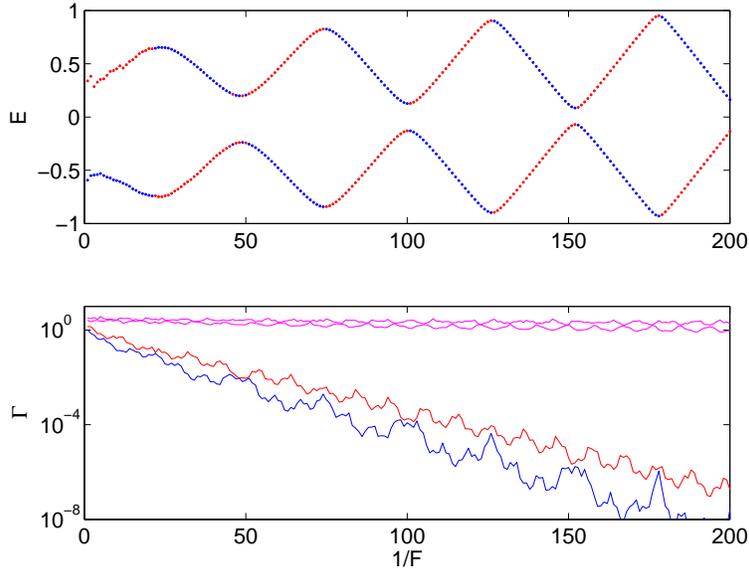}
\caption{(Color online) Position of two WS-ladders, upper panel,  and their stability (inverse lifetime), lower panel, as the function of $1/F$. In the lower panel we also depict the resonance widths $\Gamma_\alpha$ for the WS-ladders originated from the 3rd and 4th Bloch bands.}
\label{fig7}
\end{figure}

If $F\ne0$ every Bloch band  in Fig.~\ref{fig1}(b) originates a WS-ladder with equidistant spectrum. However, unlike the tight-binding model, the energy levels are now complex numbers,
\begin{equation}
\label{a17}
E'_{n,\alpha}= E_{n,\alpha}+i\Gamma_\alpha
\end{equation}
(here $\alpha$ is the band index), and the associated WS-states are metastable states (quantum resonances) with finite life-time that is inverse proportional to the resonance width $\Gamma_\alpha$ \cite{53}.  Of course, only long-living states originated from two  lower bands are of physical importance. These two ladders, reduced to the  fundamental energy interval $|E|\le F/2$, are shown in  the upper panel in Fig.~\ref{fig7}.  This figure should be compared with Fig.~\ref{fig3}   where one can see a similar structure with  progressively decreasing gaps of the avoided crossings. We note that, even  if an avoided crossing is not resolved on the scale of the figure, we can indicate its presence sorting the level  according to their stability.  In fact, if the real parts of the complex  energy levels undergo an avoided crossing, the imaginary  parts must undergo  the real crossing \cite{53}.  This behavior is clearly observed in Fig.~\ref{fig7}(b)  where WS-ladders exchange their stability at the avoided crossings.

As expected, we find the strongest deviation of the original system from its tight-binding counterpart in the limit of strong fields. In this domain coupling with higher ($\alpha>2$) Bloch bands results in non-analytic behavior of $E_{n,\pm}=E_{n,\pm}(F)$ which is seen as discontinuity of the curves in Fig.~\ref{fig7}(a). Nevertheless, the above conclusion that  two WS-ladders merge into a single ladder in the limit $F\rightarrow\infty$ remains valid.

\section{Conclusions}
\label{sec5}

We analyzed the energy spectrum of a quantum particle in a 1D double-periodic lattice in the presence of a static field $F$. It was shown that in the limit of weak fields the  spectrum consists of two Wannier-Stark ladders originated from two Bloch subbands. Each of these ladders  is proved to be uniquely characterized by two parameters -- the mean energy and geometrical  (Zak) phase of the Bloch subbands. An additional characteristic of  the spectrum is the size of the gap of the avoided crossings between energy levels  associated with two different ladders. These avoided crossings occur at certain values of $F$ and correspond to resonant interband Landau-Zenner tunneling.  As $F$ is decreased, the gap of the avoided crossings exponentially decreases.  In the opposite case, when $F$ is increased, the gaps progressively  increase and sooner or latter become comparable with the ladder step. This results in rearrangement of the spectrum from a superposition of two ladders with the step $2aF$ into a single  ladder with the step $aF$ (here $a$ is the distance between the nearest sites).  By mapping the problem to an effective two-level system we derived analytic expressions that describe this rearrangement of the spectrum. Remarkably, for one of considered in the work lattices this effective system coincides with the driven two-level system in the strong-coupling regime. Thus we demonstrated that the latter problem, which is of large importance in quantum optics,  can be viewed as a particular case of the Wannier-Stark problem for double-periodic lattices.

Finally, we analyzed the Wannier-Stark spectrum of a quantum particle in a double-periodic lattice beyond the tight-binding two-band approximation.  The above listed results were shown to hold true for the original continuous system where the Wannier-Stark states are quantum resonances and, hence, have a finite lifetime.

The authors acknowledge financial support of of Russian Academy of Sciences through the SB RAS integration project No.29 {\em Dynamics of atomic Bose-Einstein condensates in optical lattices} and the RFBR project No.15-02-00463 {\em Wannier-Stark states and Bloch oscillations of a quantum particle in a generic two-dimensional lattice}.


\newpage
\section{Appendix}
The solution of Eq.~(\ref{14a}) according to the second order Wu-Yang procedure Ref. \cite{Wu07} could be written as
\begin{equation}
\nonumber
\tilde{{\bf Y}}(t)=U(t)\tilde{{\bf Y}}(0),
\end{equation}
where $U(t)$ is the $2\times2$ matrix with elements given by
\begin{eqnarray}
\nonumber
U_{11}=e^{i\beta}\left([\cos\tau\cos\phi-i\sin\tau\sin\phi]\cos\psi - [i\cos\tau\sin\phi-\sin\tau\cos\phi]\sin\psi \right) \\
\nonumber
U_{12}=e^{-i\beta}\left([\cos\tau\cos\phi-i\sin\tau\sin\phi]\sin\psi + [i\cos\tau\sin\phi-\sin\tau\cos\phi]\cos\psi \right) \\
\nonumber
U_{21}=e^{i\beta}\left([\sin\tau\cos\phi+i\cos\tau\sin\phi]\cos\psi - [i\cos\tau\cos\phi+\sin\tau\sin\phi]\sin\psi \right) \\
\nonumber
U_{22}=e^{-i\beta}\left([\sin\tau\cos\phi+i\cos\tau\sin\phi]\sin\psi +[i\cos\tau\cos\phi + \sin\tau\sin\phi]\cos\psi \right)
\end{eqnarray}
The functions $\beta=\beta(t)$, $\tau=\tau(t)$, $\phi=\phi(t)$, and $\psi=\psi(t)$ are defined through the following equations
\begin{equation}
\nonumber
\tau(t)=\epsilon\int\limits_0^t dt'\sin[2\Omega \sin t']=\epsilon {\cal I}(t,2\Omega)
\end{equation}
\begin{equation}
\nonumber
\beta(t)=\epsilon\int\limits_0^t dt'\cos[2\Omega \sin t']\cos\left[2\epsilon {\cal I}(t',2\Omega)\right]
\end{equation}
\begin{equation}
\nonumber
\phi(t)=-\epsilon\int\limits_0^t dt'\cos[2\Omega \sin t']\sin\left[2\epsilon {\cal I}(t',2\Omega)\right]
\cos\left(2\pi\epsilon {\cal J}_0(2\Omega)-2\epsilon\int\limits_0^{t'}dt''\sin[2\Omega \cos t'']\right)
\end{equation}
\begin{equation}
\nonumber
\psi(t)=\epsilon\int\limits_0^t dt'\cos[2\Omega \sin t']\sin\left[2\epsilon{\cal I}(t',2\Omega)\right]
\sin\left(2\pi\epsilon {\cal J}_0(2\Omega)-2\epsilon\int\limits_0^{t'}dt''\sin[2\Omega \cos t'']\right)
\end{equation}

\newpage




\end{document}